\documentclass[12pt,a4paper]{article}

\usepackage[utf8x]{inputenc}
\usepackage{graphicx}
\usepackage{amsmath}
\usepackage{amssymb}
\usepackage{float}
\usepackage[font=small,labelfont=bf]{caption}
\usepackage{url}

\usepackage[sort&compress,numbers,super,comma]{natbib}
\usepackage{natmove}  

\usepackage{setspace}
\newcommand{\onlinecite}[1]{\hspace{-1 ex} \nocite{#1}\citenum{#1}}

\title{\textbf{Systematic Microsolvation Approach with a Cluster-Continuum Scheme and Conformational Sampling}}
\author{Gregor N.\ Simm, Paul L.\ T\"urtscher, and Markus Reiher\thanks{corresponding author: markus.reiher@phys.chem.ethz.ch}
\vspace{10 mm}\\
ETH Z\"urich, Laboratory of Physical Chemistry, \\ Vladimir-Prelog-Weg 2, 8093 Z\"urich, Switzerland
}
\date{11 January 2020}

\begin{document}

\maketitle

\begin{center}
\textbf{Abstract}
\end{center}

Solvation is a notoriously difficult and nagging problem for the rigorous theoretical description
of chemistry in the liquid phase. Successes and failures of various approaches ranging from
implicit solvation modeling through dielectric continuum embedding
and microsolvated quantum chemical modeling to explicit molecular dynamics highlight
this situation. Here, we focus on quantum
chemical microsolvation and discuss an explicit conformational sampling ansatz to make this approach systematic.
For this purpose, we introduce an algorithm for rolling and automated microsolvation of
solutes. Our protocol takes conformational sampling and rearrangements in the solvent shell
into account. Its reliability is assessed by monitoring the evolution of the spread and average of
the observables of interest.

\section{Introduction}

Solvation can, depending on the type of solvent into which a molecular system is immersed, strongly modulate the properties of a solute \cite{Kirchner2007,Reichardt2010}.
For example, the kinetics of reactions in solution can be affected by the solvent.
To that end, experimental studies have investigated the interactions between solute and solvent (see, for examples,
Refs.~\onlinecite{Kropman2001,Laage2007,Park2007}).
A solvent can take the role of an observer in which it assists the course of a reaction solely through non-covalent
intermolecular interactions.
It can also actively participate in a reaction in such a way that well-structured intermediates inlcuding solvent molecules are formed.
Therefore, the accurate theoretical description of chemical processes requires not only the elucidation of all relevant intermediates and elementary reactions (see Refs.~\onlinecite{Sameera2016,Dewyer2017,Simm2019})
but also adequate modeling of the reactants' environment.

There are three main approaches of including solvation effects in theoretical models:
implicit solvation, hybrid cluster-continuum schemes, and explicit solvation.
They differ in accuracy and computational cost.
In the following, the different approaches are briefly reviewed mainly in the light of their applicability to the study of chemical
reaction mechanisms.

Molecular dynamics simulations can describe a solute's dynamic surrounding at a given temperature.
Particularly suited for the solvation of reactive species are
\textit{ab initio} molecular dynamics simulations\cite{Marx2009} as they do not require the choice of
some hard-wired interaction potentials, but rely on the exactly known expression for the non-relativistic electromagnetic
interactions of elementary particles.
However, as the configuration space can become very large, computational costs of carrying out first-principles calculations grow rapidly.
As a result, they cannot be performed for every intermediate in large reaction networks\cite{Maeda2013,Zimmerman2015,Bergeler2015,Simm2017a,Dewyer2017,Simm2019}.
This issue can be overcome by the application of a reactive force-field.\cite{vanDuin2001,Dontgen2015,Silva2018} 
Unfortunately, next to the reduced accuracy, force-field parameters will, in general, not be available for any type of system which limits their applicability.
For that reason, hybrid quantum-mechanical/molecular-mechanical (QM/MM) approaches have been frequently applied
to explore complex systems with many degrees of freedom such as reactions in aqueous solution
(e.g., Ref.~\onlinecite{Galvan2004} and reviews by Senn and Thiel~\cite{Senn2007a,Senn2007,Senn2009}).
Many studies have been devoted to studying the effect of the number of explicitly treated solvent molecules
on chemical reactivity and optical properties (for examples, see Refs. \onlinecite{Neugebauer2005,Murugan2010,Flaig2012,Zuehlsdorff2016,Provorse2016,Milanese2017}).
Recently, Boereboom et al.\cite{Boereboom2018} explored multiscale approaches for the description of a reversible and highly solvent-sensitive nucleophilic bond formation reaction.
These studies demonstrate that, while the size of the solute affects the number of QM water molecules necessary to achieve convergence,
additional system-specific properties such as solvent polarity, electronic structure of the solute, and solute-solvent interactions
determine the required number of solvent molecules.

An implicit solvent model simplifies the interactions between the solute and the solvent by describing
the solvent as a polarizable medium with a solvent specific dielectric constant.\cite{Miertus1981,Tomasi2004,Tomasi2005}
The solute is then placed in a cavity formed by this medium and the interaction between the solute and solvent is calculated at the cavity boundaries.
There exist many implicit solvent models including the polarized continuum model (PCM)\cite{Miertus1981},
the conductor-like screening model (COSMO)\cite{Klamt1993}, integral equation formalism (IEFPCM)\cite{Tomasi1999}, and COSMO for 'real solvents' (COSMO-RS)\cite{Klamt1995}, and the reference interaction site model (RISM)\cite{Kovalenko2000}.
Implicit solvent models remain popular mainly for three reasons:
(i) implicit solvation is, by far, the computationally cheapest approach of modeling solvation effects,
(ii) most quantum chemistry programs allow for the activation of an implicit solvation model in a straightforward fashion, and
(iii) a continuum model can provide efficient access to free energies, even if the electrostatic contribution to the enthalpy is negligible \cite{Klamt1995,Ben-Naim,Truhlar1,Truhlar2}.

However, implicit solvent models fall short in many practical cases.
First, they are known to describe strong (directed) interactions between solute and solvent (e.g., hydrogen bonds) poorly.
This is often the case when the solute is charged (or contains a charged moiety) and the solvent is polar.
Second, they will fail if solvent molecules can react with the solute.
This includes cases in which intermediates are formed, protons are shuttled, or functional groups
such as aldehydes and ketones undergo tautomerization.

In 2015, Plata and Singleton's study of the Morita-Baylis-Hillman reaction highlighted issues of implicit solvation models for this specific reaction.\cite{Plata2015}
Recent work on this 
reaction by Basdogan and Keith\cite{Basdogan2018}
demonstrated the necessity of explicit solvation when studying organic reaction mechanisms.
However, the importance of explicit solvation has been shown numerous times for chemical reactions (see, e.g., Refs.~\onlinecite{Park2006,Michel2011,Glaves2012,DeWispelaere2016,Boereboom2018}).
Explicit solvation is also critical for physico-chemical properties (such as pK$_\text{a}$
values of organic compounds\cite{Thapa2017}), spectroscopic properties (such as NMR chemical shifts\cite{Roggatz2018,Caputo2018}),
and absorption spectra\cite{Neugebauer2005,Jacob2006,Georg2007,Neugebauer2010,Milanese2017}.

A popular attempt to remedy the shortcomings of implicit solvent models is the
introduction of explicit solvent molecules to the system.\cite{Pliego2001,Kelly2006,Coutinho2007,Bryantsev2008,Kua2013,Kua2013a}
The goal of such hybrid cluster-continuum schemes is to model short-ranged interactions explicitly
and long-range effects through the continuum model surrounding the cluster (for a recent review see Ref.~\onlinecite{Pliego2019}).
However, unless one knows the solute's local solvent environment \textit{a priori}, this approach has multiple pitfalls:
First, it is unclear how many solvent molecules need to be added to describe solvent effects to the desired accuracy.
Second, the manual process of adding solvent molecules to selected regions around a solute molecule is often guided by
ad hoc assumptions.
Third, due to the unfavorable scaling of most quantum chemical methods often only one (rarely a few)
low-lying solute-solvent configurations are taken into consideration.
In the light of the high dimensionality and rugged nature of the potential energy surface (PES) of solvent-solute clusters, it is
unlikely to find a representative configuration through manual exploration.

Recent studies attempted to tackle the above-mentioned issues of the static quantum chemical approach.
In an extended study of solute-solvent complexes that applied global structure optimization techniques, Li and Hartke\cite{Li2013}
explored extreme effects of solvent molecules on chemical reactions (in particular, on barrier heights) that will hardly be
seen if one departs from the most likely distribution of solvent molecules around the solute. However, without a measure
that informs about the probability of forming such solute-solvent structures, it remains uncertain whether extreme
cases are likely to play a role in a reaction in solution. It is therefore mandatory to provide some way
of configurational sampling to acquire information on the density of structures in a relevant energy range.
In their work on the Morita-Baylis-Hillman reaction, Basdogan and Keith\cite{Basdogan2018}
found that a stochastic computational filtering procedure using a global optimization code can help identify low energy structures.
They justified the number of solvent molecules in the solvation model by comparing computational results with experiment.
In this way, they favored a model consisting of five methanol molecules over a model with ten.
Kildgaard et al.\cite{Kildgaard2018,Kildgaard2018a} presented a stochastic algorithm for the generation of hydration clusters of sulfuric acid.
Their algorithm places water molecules in selected orientations around a solute in an iterative fashion.
However, they have not attempted to obtain a distribution over conformations but only the conformation with the lowest free energy was sought.
Moreover, the algorithm is not applicable to any choice of solvent but has been tailored for the construction of water clusters.

Considering the shortcomings of current approaches, a method is sought that fulfills the following key requirements:
\begin{enumerate}
  \item\label{req:feas} \textbf{Computational Feasibility}: a cost-effective model is required that strikes the right balance between accuracy and computational feasibility.
  \item\label{req:improv} \textbf{Systematic Improvability}: the accuracy of the model should be adaptable to the computational resources available
  and, in principle, be systematically improvable so that the physically correct description of solvation in the proper thermodynamic ensemble is recovered.
  \item\label{req:univ} \textbf{Universal Applicability}: the model should be applicable to any solute-solvent combination and even solvent mixtures.
  \item\label{req:auto} \textbf{Full Automation}: the approach should not require any human intervention to be unbiased and to avoid wasting human time.
\end{enumerate}

In this paper, we present a static cluster-continuum scheme that fulfills these requirements.
Regarding the first requirement, we apply efficient generalized-gradient-approximation density functional theory with density fitting for
the optimization of solvent-solute clusters, but note that
semi-empirical quantum chemical methods\cite{Husch2018a} can very much reduce the increasing computational cost caused by
considering solvent molecules explicitly.\cite{Basdogan2018,Kildgaard2018a}
As these methods struggle to properly describe important interactions in solution such as dispersion and hydrogen-bonding,
corrections have been developed to alleviate such shortcomings\cite{Korth2010,Korth2010a,Korth2011}. Hence,
their application for the conformational sampling of large clusters will be beneficial.
We develop our approach at the example of a prototypical system: acetonitrile in the solvents water and dichloromethane (DCM).
The corresponding algorithm has been integrated into our \texttt{Chemoton} structure exploration program\cite{Simm2017a} that is
part of the \texttt{SCINE} software suite.\cite{scineWeb}

\section{Methodology}

\subsection{Stochastic Generation of Solvation Clusters}
\label{sec:protocol}

The algorithm presented below ensures that solvent molecules are placed equally around any solute in a stochastic fashion.
To ensure that requirements~\ref{req:univ} 'Universal Applicability' and \ref{req:auto} 'Full Automation' are fulfilled,
the generation of solvation clusters may only depend on nuclear coordinates and quantum chemical observables such as the electron density.

\begin{figure}[!htb]
\begin{center}
\includegraphics[width=0.6\textwidth]{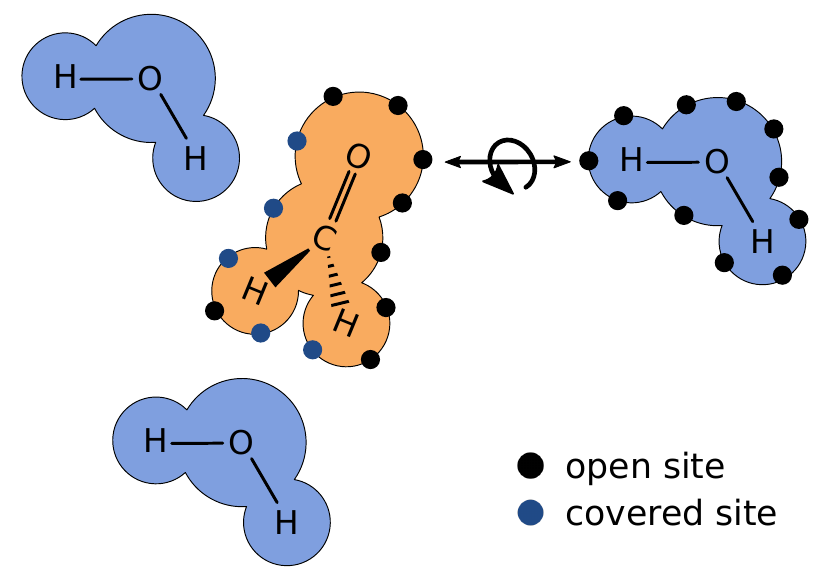}
\caption{
Open (black) and covered (blue) sites on the accessible surface of solute (formaldehyde) and solvent (water).
Sites on the added solvent molecules are not considered until all sites at the solute are covered.
This ensures that solvent molecules are equally distributed around the solute.
}
\label{fig:sites}
\end{center}
\end{figure}

\begin{enumerate}
  \item\label{alg:start} First, the accessible surfaces of the solute-solvent complex (which consists of the solute only in the beginning) and solvent are identified.
  For a given molecular structure, an icosahedral mesh consisting of 12 vertices ${s_a}$ (called sites) is created around each atom $a$ with radius equal to its van der Waals radius $R_a$.
  With mesh sub-division algorithms, meshes of arbitrary smoothness can be generated.
  \item\label{alg:sites} Sites will be marked as covered if a ray originating from the site going in the direction of the site's surface normal hits another mesh within a distance of $d_\text{cutoff}$.
  All sites that are buried by meshes from neighboring atoms are removed.
  The concept of (\textit{open} and \textit{covered}) sites is illustrated in Fig.~\ref{fig:sites}.
  \item\label{alg:select} Two open sites $s_{a,\text{solute}}$ and $s_{b,\text{solvent}}$ of the solute-solvent complex and the solvent, respectively, are selected at random.
  \item If there were no open sites in the solute, i.e., all sites are covered by the solute itself or by solvent molecules, a complete solvation shell had been formed.
  If more solvent molecules are to be added, the solute-solvent complex will be considered the new starting cluster and the algorithm continues with step~\ref{alg:start}.
  \item The solute-solvent complex and the solvent molecule are arranged relative to each other in such a way that site $s_{a,\text{solute}}$, atom $a$, site $s_{b,\text{solvent}}$, and atom $b$ are on one axis.
  The distance between the sites $s_{a,\text{solute}}$ and $s_{b,\text{solvent}}$ is set to $d$.
  \item\label{alg:angle} The angle around this axis is chosen randomly under that constraint that atoms do not come too close to each other, i.e., their van der Waals spheres do not overlap.
  \item If no orientation can be found for which atoms do not come too close to one another,
  $d$ will be increased in increments of $d_\text{inc}$ and step~\ref{alg:angle} will be attempted again until a maximum distance $d_\text{max}$ will be reached.
  \item If it is not possible to place a solvent molecule, the site will be marked as covered and the algorithm continues with step~\ref{alg:select}.
  \item Once a solvent molecule is added to the solute-solvent complex, the algorithm continues with step~\ref{alg:sites}
  until the desired number of solvent molecules has been added.
\end{enumerate}

The random selection of sites in step~\ref{alg:select} will not be optimal if the solute or solvent features charged
polar functional groups or those containing strong H-bond donors or acceptors.
Depending on the polarity and charge of the solute and solvent an equal distribution of solvent molecules around the solute may be inefficient.
To take this into account, one could augment sites with descriptors based on first principles such as the Laplacian of the electron density,\cite{Bader1994}
Fukui functions,\cite{Fukui1982} partial atomic charges,\cite{Mulliken1955,Mulliken1955a,Meister1994,Bultinck2007} atomic polarizabilities,\cite{Brown1982,Kang1982,Kunz1996}
or dual descriptors\cite{Morell2005,Morell2006,Ayers2007,Cardenas2009}
(see also Refs.~\onlinecite{Geerlings2003,Geerlings2008,Johnson2011} for reviews).
As a result, electron-rich oxygen atoms, for example, would be placed in such a way that they are facing electron-deficient hydrogen atoms.
Such an approach could reduce the number of steps required in the structure optimization and improve the probability of generating a low-energy configuration.
At the same time, care should be taken so that the use of descriptors does not introduce any bias.

If the solvent has multiple relevant conformations (e.g., hexane) or a mixture of solvents is to be modeled,
multiple structures (possibly with corresponding statistical weights) can be included in the site selection process in step~\ref{alg:select}.
Although these extensions are straightforward to implement, they are beyond the scope of the present work.

\subsection{Hybrid Solvation Model}

\begin{figure}[!htb]
\begin{center}
\includegraphics[width=0.8\textwidth]{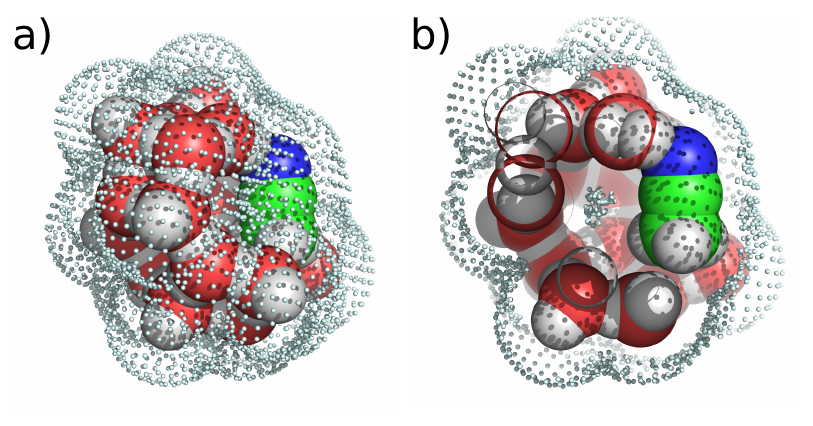}
\caption{
Acetonitrile-water cluster surrounded by a dielectric continuum (light blue spheres).
A cross-section (b) reveals an unphysical cavity within the cluster.
}
\label{fig:leakage}
\end{center}
\end{figure}

The generated solute-solvent complexes are embedded in a cavity formed by a continuum with the dielectric constant of the solvent $\epsilon_\text{solv}$\cite{Pye1999}.
For the construction of the cavity, its solvent accessible surface needs to be determined.
This can be achieved by probing the complex with a sphere of radius $R_\text{solv}$\cite{Pye1999} that is specific to the solvent.
This procedure was originally developed for single molecules, not for molecular clusters.
In the latter case, unphysical cavities can form within the complex (see Fig.~\ref{fig:leakage}).
In practice, however, these cavities can be easily detected and a larger $R_\text{solv}$ can be chosen to prevent their formation.
The small deviation from the idealized radius has a negligible effect on the solute-solvent interaction,
especially as the border of the cavity should be far from the solute.

The resulting cluster structure is then subjected to quantum chemical structure optimization that
will lock it into a local minimum on the Born-Oppenheimer PES.
One could, however, also optimize the structures after each addition of a solvent molecule.

\subsection{Sampling of Solvation Clusters and Convergence}
\label{sec:sampling}

In the canonical ensemble, the importance of a set of nuclear coordinates $\mathbf{x}$ (the configuration)
of a molecular system at temperature $T$ may be taken as governed by a Boltzmann probability distribution,
\begin{equation}
\label{eq:boltz}
  p(\mathbf{x}) = \frac{1}{z} \exp \left\{- E(\mathbf{x}) /(k_\text{B} T) \right\},
\end{equation}
with the total energy of the configuration $E(\mathbf{x})$,
the molecular contribution $z$ to the canonical partition function, and the Boltzmann constant $k_\text{B}$.
For the sake of simplicity, we assume that the total energy is governed by the electronic energy of each solute-solvent
cluster in a dielectric continuum so that we can rate the different configurations of one cluster size according to this quantity
(nuclear zero-point and finite-temperature contributions may be added through the standard translation-in-a-box harmonic-oscillator rigid-rotor approximation for
the partition function equipped with a dielectric-continuum solvation model for assessing the change in free energy
associated with embedding a solute-solvent cluster into the liquid phase).

$M$ samples drawn from this distribution then yield a configuration average of an observable $\langle O \rangle$,
\begin{equation}\label{eq:int}
  \langle O \rangle = \frac{1}{M} \sum_{i=1}^M O(\mathbf{x}_i) .
\end{equation}
A strategy now needs to be devised which
ensures that sufficiently many \textit{relevant} configurations are generated so that
the configurations are assigned the proper weight $p(\mathbf{x}_i)$ when calculating expectation values $\langle O \rangle$
for observables $O$,
\begin{equation}\label{eq:sum}
  \langle O \rangle = \sum_{i=1}^M p(\mathbf{x}_i) O(\mathbf{x}_i) ,
\end{equation}
with the weights normalized according to Eq.~(\ref{eq:boltz}) by
\begin{equation}
\label{eq:z}
z\approx  \sum_{j=1}^M \exp \left\{- E(\mathbf{x}_j) /(k_\text{B} T) \right\}.
\end{equation}
This is non-trivial (especially for systems with many floppy degrees of freedom such as solute-solvent clusters). For this reason,
the inclusion of solvent molecules is often avoided or only the lowest-total-energy configuration $\mathbf{x}_\text{minimum}$ is considered,
\begin{equation}
  \langle O \rangle \approx O(\mathbf{x}_\text{minimum}).
\end{equation}
Such an approximation of $\langle O \rangle$ by a single sample will only be reasonable if it can be guaranteed
that there exists a sufficiently large energy gap between this single configuration and all others higher in energy.
In general, this assumption will not be justified for rugged PESs such as those created by explicit solvation.

To ensure that our approach is systematically improvable (requirement~\ref{req:improv}) yet computationally feasible (requirement~\ref{req:feas}),
we present a procedure that allows one to
a) systematically determine a minimum number of solvent molecules required for modeling the solvent effect to sufficient
accuracy and b) approximate the distribution over configurations of solute-solvent clusters to obtain reliable ensemble
properties at a given temperature.

\begin{figure}[!htb]
\begin{center}
\includegraphics[width=0.9\textwidth]{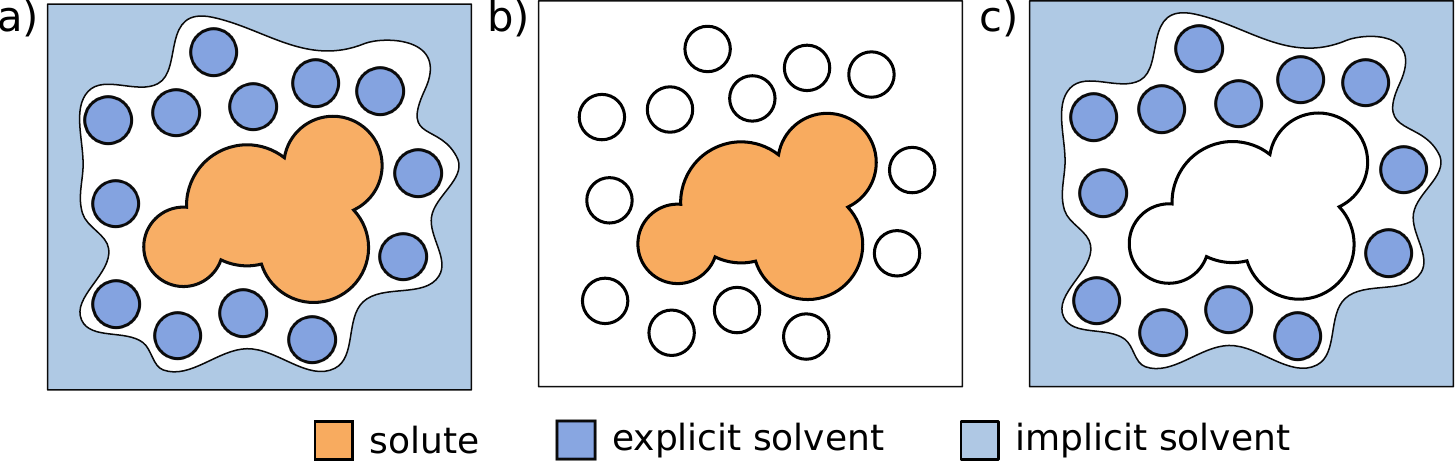}
\caption{
Illustration of the components of $E_\text{inter}$ in our hybrid cluster-continuum solvation scheme:
(a) solute, solvent molecules, and continuum model ($E_\text{total}$),
(b) solute ($E_\text{solute}$),
(c) solvent molecules and continuum model surrounding empty cavity formed by solute ($E_\text{solvent}$).
To avoid basis set superposition errors, empty nuclear positions are occupied by ghost atoms.
}
\label{fig:energies}
\end{center}
\end{figure}

We now define the (electronic) interaction energy $E_\text{inter}$ between the solute and its environment in the following way:
\begin{equation}\label{eq:inter}
  E_\text{inter} = E_\text{total} - E_\text{solute} - E_\text{solvent},
\end{equation}
where $E_\text{total}$ is the total electronic energy of the solute-solvent complex,
$E_\text{solute}$ is the electronic energy of the solute,
and $E_\text{solvent}$ is the electronic energy of the solvent molecules including the implicit solvent model
(see Fig. \ref{fig:energies} for a graphical representation of these energy components).

If the total electronic energy $E_\text{total}$ of the $i$-th cluster had been used in
Eqs.~(\ref{eq:sum}) and (\ref{eq:z}), it would be strongly dominated by configurations in which the solvent molecules (most of which will be far away from the solute)
are arranged in a stable configuration.
This effect would draw the attention from the solute to the solvent, particularly in the case of large clusters.
As a remedy, we employ the following expression for the energy in the Boltzmann weighting in Eq.~(\ref{eq:boltz}):
\begin{equation}\label{eq:solute_env}
  E_\text{solute+env} = E_\text{solute} + E_\text{inter} = E_\text{total} - E_\text{solvent}
\end{equation}
In this way, also the interaction between the solvent molecules and the continuum model are quenched
(this assumes that the solute is not in direct contact with the continuum).
With this partitioning scheme, however, free energies cannot be straightforwardly calculated
because vibrational modes cannot (easily) be attributed to the solute and its direct surrounding
(see also the recent comparison of finite-temperature models for solution in Ref.~\onlinecite{Besora2018} and references therein).
We note that a recent study\cite{Ho2015} showed that the inclusion of vibrational contributions in the free energy
of solvation can have a negligible effect on the accuracy of thermodynamic cycle predictions of pK$_\text{a}$'s and reduction potentials.

The complete algorithm then runs as follows:
\begin{enumerate}
  \item\label{alg:construct} Generate $N_\text{sample}$ solute-solvent complexes by adding $n_\text{solv}$ solvent molecules to the solute
  with the procedure detailed in Section~\ref{sec:protocol}.
  \item Optimize the structures of the complexes.
  \item Evaluate $\langle E_{\text{inter}} \rangle$ and $\langle E_\text{distort} \rangle$ employing $E_\text{solute+env}$ for the weights $w$.
  $E_\text{distort}$ is defined as the energy of the solute (without counterpoise correction) from which the energy of the solute optimized in vacuum
  is subtracted.
  With $\langle E_{\text{inter}} \rangle$ and $\langle E_\text{distort} \rangle$ we have measures quantifying
  the solute interactions with the solvent and the distortion of the solute induced by the environment, respectively.
  If the cluster partition function $z$ is dominated by only a few configurations, more samples must be generated in step~\ref{alg:construct}.
  \item If $\langle E_{\text{inter}} \rangle$ and $\langle E_\text{distort} \rangle$ are sufficiently close to the values of a set of clusters consisting of fewer solvent molecules, terminate.
  Else, increase $n_\text{solv}$ and go back to step~\ref{alg:construct}.
\end{enumerate}

This algorithm is based on first principles and does not depend on the solute's shape or charge.
As a result, it will, in principle, be applicable to any solute, also to those that are strongly interacting with the solvent.
In these cases, more solvent molecules may be required until the convergence criteria are met.

Further, when a larger complex is generated, its structure is not based on a cluster from the previous iteration to ensure that no bias is introduced.

\section{Computational Details}

The protocol described so far was implemented into our program package \texttt{Chemoton}\cite{Simm2017a}
and carried out in a fully automated fashion.
Icosahedral meshes were generated with the C++ library OpenMesh\cite{openmesh71Web}.
We will make \texttt{Chemoton} including the features described in this work available through our SCINE web page.\cite{scineWeb}

All structure optimizations and single-point calculations were carried out with
the \texttt{Orca} program package (version 4.0.1)\cite{Neese2012}
employing the exchange-correlation density functional PBE\cite{Perdew1996a},
D3BJ dispersion corrections\cite{Grimme2010,Grimme2011},
the def2-TZVPP\cite{Weigend2005,Weigend2006} basis set,
and the corresponding density-fitting resolution-of-the-identity approximation for the Coulomb integrals.
For structure optimizations and single-point calculations the default convergence criteria were chosen.
We note that for the raw-data generation other quantum chemistry packages can be easily interfaced.

For the conformer generation according to the protocol described above the following parameters were employed:
$d_\text{cutoff} = 5$~\AA,
the initial setting of $d = 0$~\AA,
$d_\text{inc} = 0.25$~\AA,
$d_\text{max} = 5$~\AA,
and $N_\text{sample} = 100$.
Between each iteration, in the protocol specified in Section~\ref{sec:sampling}, we increased the number of solvent molecules by 5.
This choice was made arbitrarily for this work, but can, in principle, be treated as a solute-size-dependent parameter

The implicit solvent was described with the conductor-like polarizable continuum solvation model (CPCM)\cite{Klamt1993,Andzelm1995,Barone1998,Cossi2003}.
The solvent probe radius was increased from 1.3 to 1.9~\AA{} to avoid cavities being formed within the solute-solvent complex.
To verify that even for small cluster sizes this parameter change does not significantly distort our results we studied an extreme case.
We ran single point calculations for acetonitrile in a continuum (i.e., no explicit solvent molecules) with two solvent probe radii: 1.3 (default) and 1.9~\AA.
With the computational setup described in the paper the electronic energy does not change significantly ($< 10^{-6}$ a.u.) for both solvents.
In addition, visual inspection of the cavities showed that they are practically the same.

When calculating the Boltzmann weight of each solute-solvent complex, the temperature was set to $T$=291.15~K.
For the calculation of $E_\text{solute}$ and $E_\text{solvent}$, a basis set superposition error was avoided
by attaching the appropriate ghost-basis functions to unoccupied nuclear positions in the energy evaluation.

The RDFs were calculated with a bin size of 1.8~a.u.
A cluster's average solvent density was estimated by dividing the number of solvent molecules by the volume of  the convex hull spanned by all atoms.

All results were saved to a Mongo database.\cite{MongoDB32}
Automated data analysis was performed with the Python libraries \texttt{matplotlib}\cite{Hunter2007}
and \texttt{pandas}.\cite{McKinney2010}
The structures obtained are provided in the Supporting Information.

\section{Results}

We study our protocol at the example of acetonitrile as solute in the solvents DCM and water.
We chose these components because
neither acetonitrile nor either solvent contains conformational degrees of freedom to be sampled in addition to the configurational ones.
Moreover, acetonitrile is soluble but chemically inert in both solvents and
contains a polar, hydrogen-bond accepting nitrile group as well as a non-polar methyl group.
Our protocol will terminate if the absolute differences with respect to $\langle E_{\text{inter}} \rangle$ and $\langle E_\text{distort} \rangle$
between the current cluster and the previous one are both below 0.5~kcal/mol (or $\approx$ 2.1~kJ/mol).
This threshold value may be adjusted as it depends on the system's size and the required accuracy.

\subsection{Acetonitrile-dichloromethane clusters}

In Table~\ref{tab:dcm}, the number of successfully optimized acetonitrile-DCM clusters $N_\text{clusters}$ (out of
$N_\text{sample} = 100$ generated starting structures)
and the number of clusters with significant contributions to the partition function $n_\text{sig}$ ($p(\mathbf{x}) > 0.05$, where $M =N_\text{clusters}$)
are given for different cluster sizes.
A structure optimization of a cluster is considered successful if the converge criteria are met within 1200 structure optimization steps.

\begin{table}[!htb]
\centering
\caption{
Number of successfully optimized clusters $N_\text{clusters}$ and number of significant clusters $n_\text{sig}$
($p(\mathbf{x}) > 0.05$) for clusters containing $n_\text{solv}$ DCM molecules.
T is the average wall time (in hours, on a 16-core CPU) and $n_\text{cycles}$ is the average number of required optimization cycles.
}
\label{tab:dcm}
\begin{tabular}{lrrrr}
\hline
\hline
$n_\text{solv}$ &  $N_\text{clusters}$ &  $n_\text{sig}$ & T & $n_\text{cycles}$\\
\hline
5            &          98 &      5 & 2  & 225  \\
10           &          99 &      4 & 8  & 264  \\
15           &          95 &      6 & 18 & 254  \\
20           &          93 &      4 & 35 & 248  \\
25           &          83 &      5 & 45 & 179  \\
\hline
\hline
\end{tabular}
\end{table}

We note that the algorithm from Section~\ref{sec:sampling} would have terminated after 15 solvent molecules.
First, it can be seen that almost all generated clusters could be optimized successfully.
This demonstrates that our algorithm produces starting structures 
that reproduce proper bonding patterns so that
subsequent structure optimizations do not yield dissociating structures such
as hydronium ions.

Second, Table~\ref{tab:dcm} shows that the number of structure optimizations that failed to converge within 1200 steps
increases with the number of solvent molecules.
The slow convergence of these structure optimizations can be attributed to the system size and its many soft degrees of freedom.
Further, the ratio between the number of significant clusters $n_\text{sig}$ and the total number of generated clusters
$N_\text{clusters}$ is relatively low ($\approx 2-5\%$) for all clusters sizes.
As described in Section~\ref{sec:sampling}, the weight of each cluster is calculated from $E_\text{solute+env}$.
Since the variance in $E_\text{solute}$ is small (see below), it is the interaction energy $E_\text{inter}$ that is responsible for the large spread.
Therefore, there must be a few configurations that have significantly higher $E_\text{inter}$ than others.

\begin{figure}[!htb]
\begin{center}
\includegraphics[width=\textwidth]{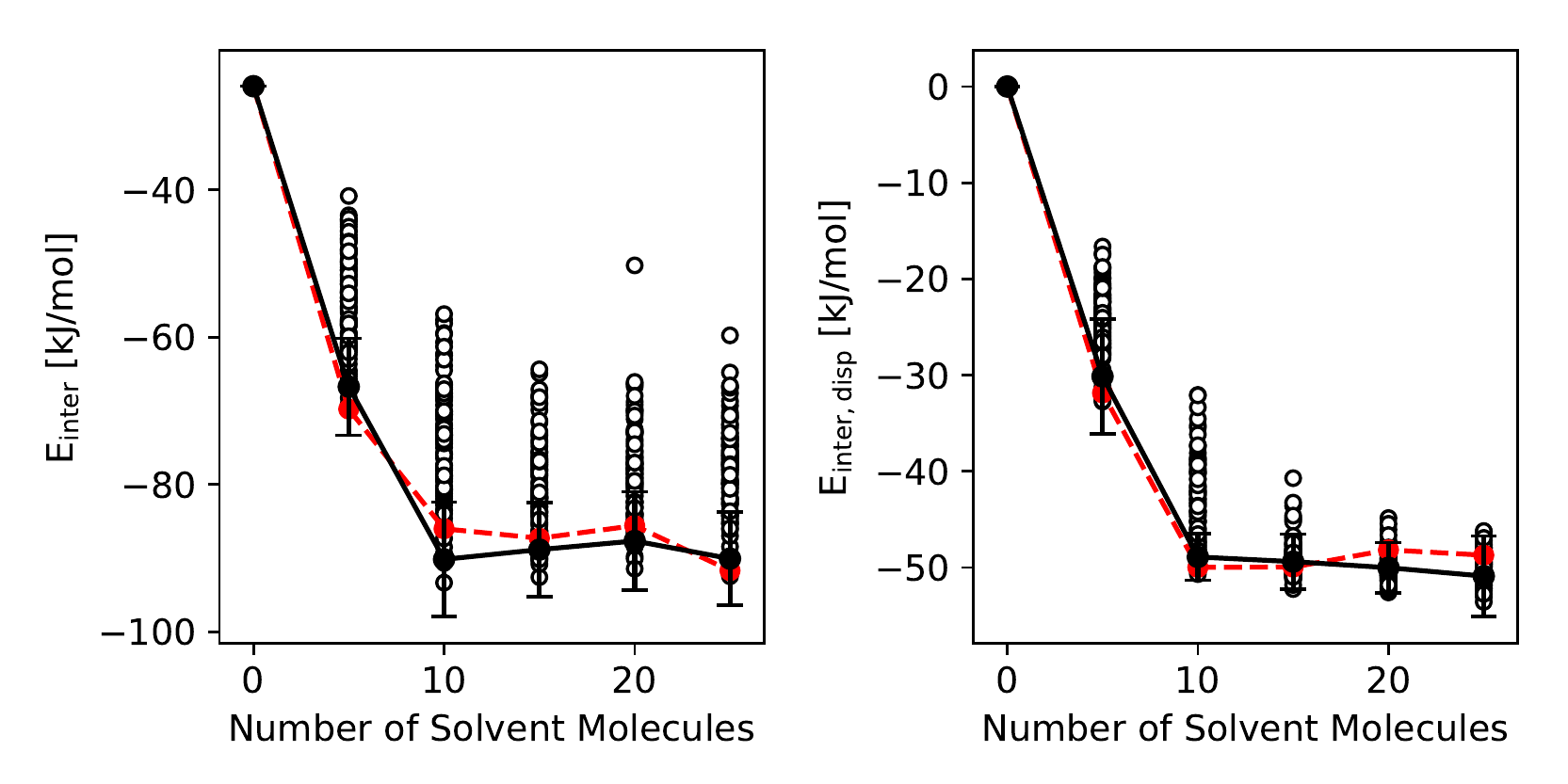}
\caption{
$E_\text{inter}$ (left) and its dispersive component $E_\text{inter, disp}$ (right) of acetonitrile-DCM complexes (white discs)
as a function of the number of solvent molecules in kJ/mol.
Black discs connected by black solid lines show $\langle E_{\text{inter}} \rangle$ and $\langle E_{\text{inter, disp}} \rangle$
(with error bars indicating $\pm 2 \sigma$).
Red discs connected by red dashed lines highlight $E_{\text{inter}}$ and $E_{\text{inter, disp}}$ of the most stable complexes.
}
\label{fig:acet_inter_dcm}
\end{center}
\end{figure}

\begin{figure}[!htb]
\begin{center}
\includegraphics[width=0.5\textwidth]{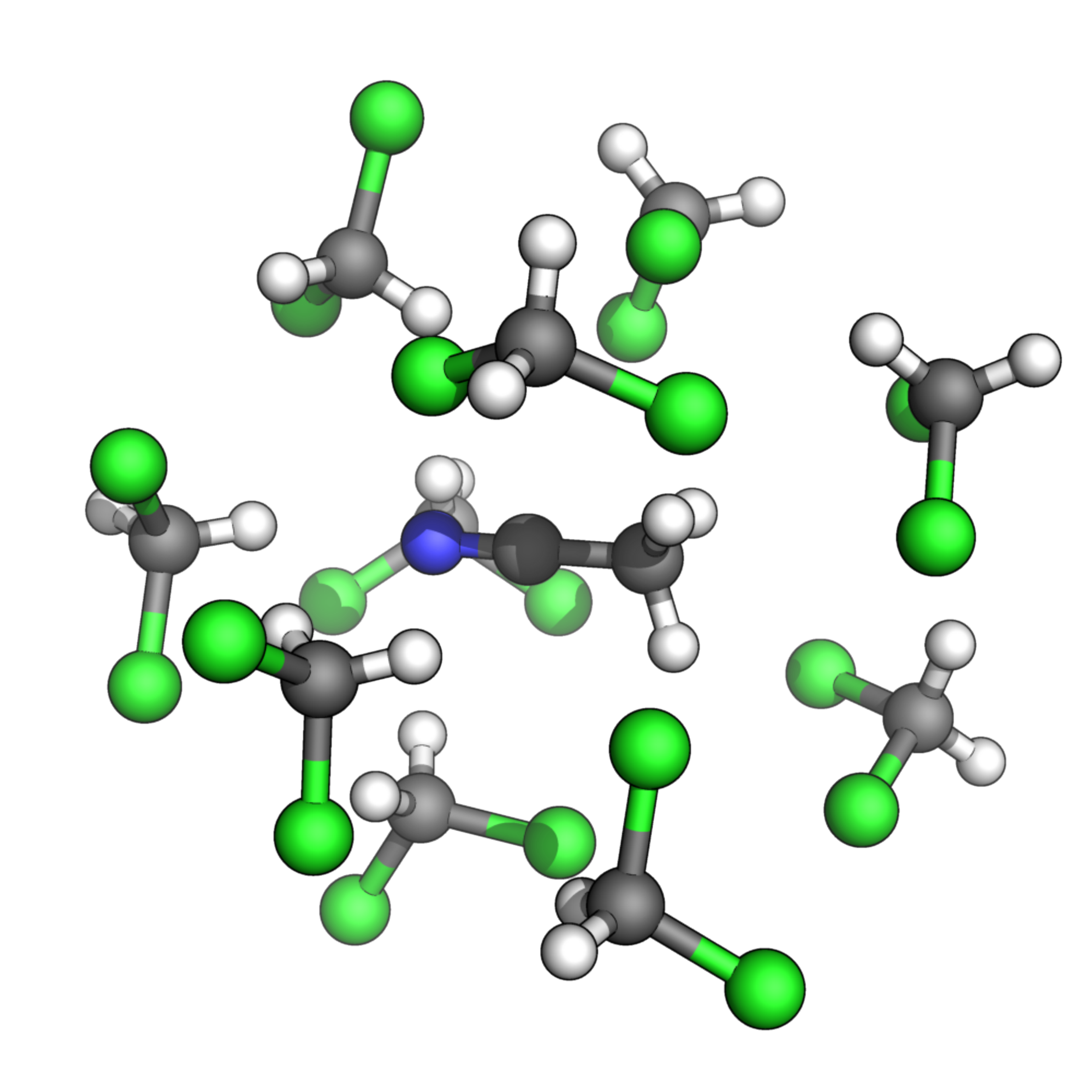}
\caption{
Ten DCM molecules form a complete solvation shell around one acetonitrile molecule.
}
\label{fig:dcm_shell}
\end{center}
\end{figure}

In Fig.~\ref{fig:acet_inter_dcm}, $E_\text{inter}$ (left) and the dispersive component of $E_\text{inter}$ (right) of acetonitrile-DCM complexes (white discs) are plotted as a function of the number of solvent molecules.
First, it can be seen that $E_\text{inter}$ drops from $-26$ to $ -90$~kJ/mol with increasing cluster size until clusters consist of ten DCM molecules.
After that point, $E_\text{inter}$ remains constant (within error bars).
A similar trend can be observed for $E_\text{inter, disp}$.
Visual inspection of the complexes shows that with ten DCM molecules a complete solvation shell can be formed (see the most stable configuration in Fig.~\ref{fig:dcm_shell}).
This may explain why the interaction energy does not change after that point:
once a full solvation shell is formed, hardly any further pronounced interactions between the solute and the additional solvent
molecules emerge for this solute-solvent combination.

Further, Fig.~\ref{fig:acet_inter_dcm} shows that $E_{\text{inter}}$ and $E_{\text{inter, disp}}$
of the most stable complexes are located close to the clusters with the strongest solute-solvent interaction (lowest $E_{\text{inter}}$ and $E_{\text{inter, disp}}$).
From this finding, we conclude that the stability of the clusters can be attributed to the interaction between solute and solvent,
rather than from interactions between solvent molecules.

\begin{figure}[!htb]
\begin{center}
\includegraphics[width=\textwidth]{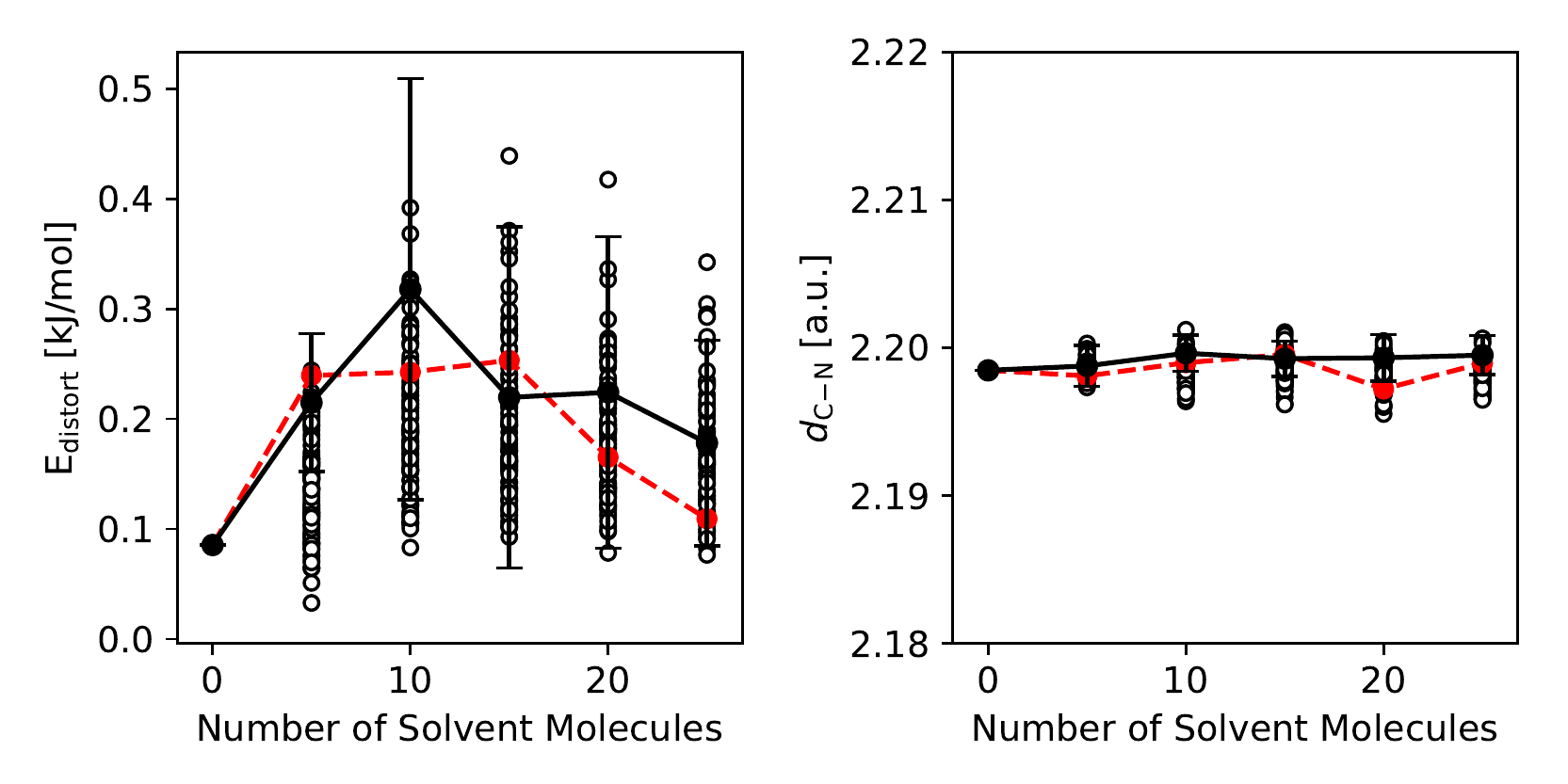}
\caption{
$E_\text{distort}$ (left, in kJ/mol) and $d_\text{C-N}$ (right, in atomic units (a.u.), i.e. bohr in this case)
of acetonitrile-DCM complexes (white discs) as a function of the number of solvent molecules.
Black discs connected by black solid lines show $\langle E_{\text{distort}} \rangle$ and $\langle d_{C-N} \rangle$
(with error bars indicating $\pm 2 \sigma$).
Red discs connected by red dashed lines show $E_{\text{distort}}$ and $d_\text{C-N}$ of the most stable complexes.
}
\label{fig:acet_solute_dcm}
\end{center}
\end{figure}

To study the effect of the environment, we also track the length of the C-N bond in acetonitrile, $d_\text{C-N}$.
In Fig.~\ref{fig:acet_solute_dcm}, $E_\text{distort}$ and $d_\text{C-N}$ are shown as a function of the number of DCM molecules in the solute-solvent complex.
It can be seen, that $E_\text{distort}$, as well as changes in $d_\text{C-N}$, are relatively small for all cluster sizes.
In addition, the spread of $E_\text{distort}$ and $d_\text{C-N}$ among clusters of the same size is small ($\approx 0.4$~kJ/mol and $\approx 0.005$ bohr, respectively).
Therefore, in this system, the main contributor to deviations in $E_\text{solute+env}$ among clusters is the interaction energy $E_\text{inter}$, not the stability of the solute.
For large molecules that can undergo large conformational changes in solution, this will not necessarily be the case.
Furthermore, it can be seen that $\langle E_{\text{distort}} \rangle$ converges (within error bars) whereas $E_{\text{distort}}$ of the most stable configuration does not.
This showcases the issue of employing the total energy as a measure for computing Boltzmann averages:
with an increasing number of explicit solvent molecules, the stability of the environment will dominate over that of the solute.

\subsection{Acetonitrile-water clusters}

\begin{figure}[!htb]
\begin{center}
\includegraphics[width=0.5\textwidth]{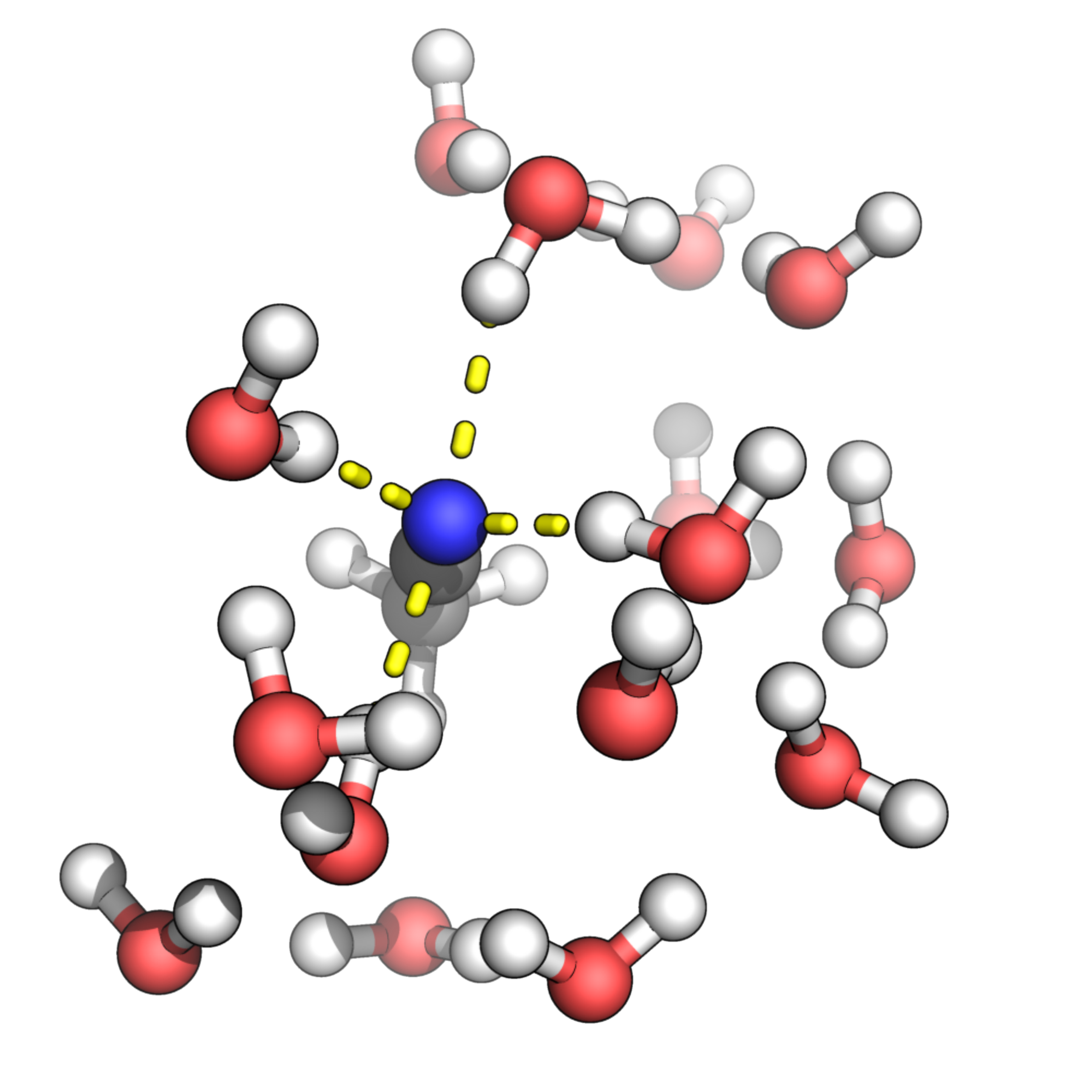}
\caption{
Formation of four hydrogen bonds (indicated by yellow dashed lines) between water and the nitrile group of acetonitrile.
Among the clusters containing 15 water molecules, this structure has the highest (absolute value of the)
interaction energy $E_\text{inter}$.
}
\label{fig:hbonds}
\end{center}
\end{figure}

As a difficult case for microsolvation, we now turn to water as the medium for acetonitrile.
In Table~\ref{tab:h2o}, the number of successfully optimized acetonitrile-water clusters $n_\text{clusters}$
and the number of clusters with significant contributions to the partition function $n_\text{sig}$ ($p(\mathbf{x}) > 0.05$, where $M = N_\text{clusters}$)
are given for different cluster sizes.
We note that the algorithm from Section~\ref{sec:sampling} would have terminated after 35 solvent molecules.
Similar trends as in Table~\ref{tab:dcm} can be identified, however, $n_\text{sig}$ is generally lower in water than in DCM.
This can be explained by the strong hydrogen-bond network formed around the nitrile group which consists of a specific arrangement of water molecules (see Fig.~\ref{fig:hbonds}).
Out of the optimized clusters, only a small fraction features this particular arrangement.

\begin{table}[!htb]
\centering
\caption{
Number of successfully optimized clusters $n_\text{clusters}$ and number of significant clusters $n_\text{sig}$
($p(\mathbf{x}) > 0.05$) for clusters containing $n_\text{solv}$ water molecules.
T is the average wall time (in hours, on a 16-core CPU) and $n_\text{cycles}$ is the average number of required optimization cycles.
}
\label{tab:h2o}
\begin{tabular}{lrrrr}
\hline
\hline
$n_\text{solv}$ &  $N_\text{clusters}$ &  $n_\text{sig}$ & T & $n_\text{cycles}$\\
\hline
5            &          97 &      7 &   1   &  230 \\
10           &          98 &      3 &   3   &  302 \\
15           &          98 &      1 &   4   &  296 \\
20           &          99 &      1 &   7   &  319 \\
25           &          99 &      3 &   12  &  328 \\
30           &          98 &      3 &   16  &  309 \\
35           &          96 &      2 &   24  &  324 \\
\hline
\hline
\end{tabular}
\end{table}

\begin{figure}[!htb]
\begin{center}
\includegraphics[width=\textwidth]{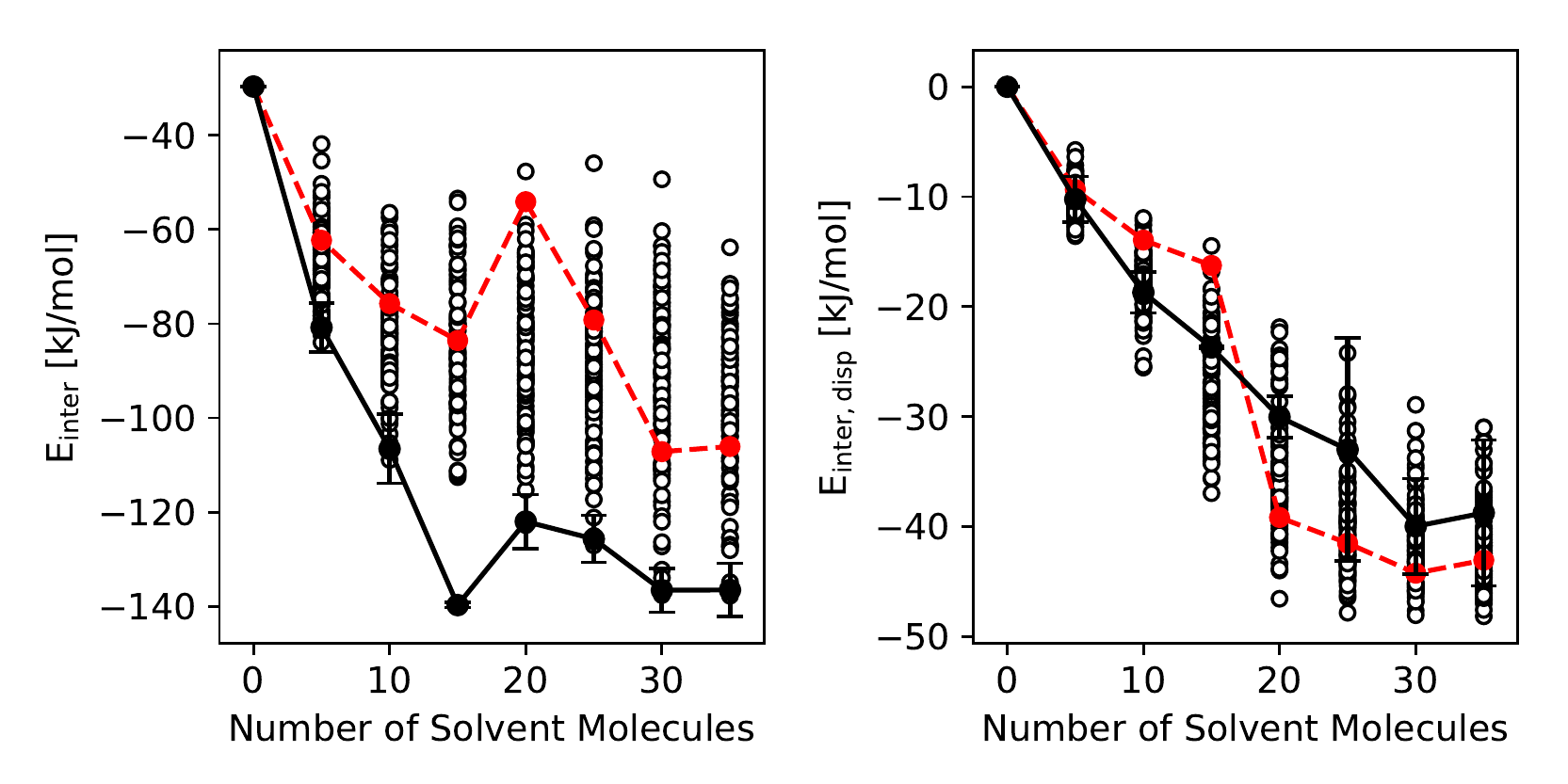}
\caption{
$E_\text{inter}$ (left) and its dispersive component $E_\text{inter, disp}$ (right) of acetonitrile-water complexes (white discs)
as a function of the number of solvent molecules in kJ/mol.
Black discs connected by black solid lines show $\langle E_{\text{inter}} \rangle$ and $\langle E_{\text{inter, disp}} \rangle$
(with error bars indicating $\pm 2 \sigma$).
Red discs connected by red dashed lines highlight $E_{\text{inter}}$ and $E_{\text{inter, disp}}$ of the most stable complexes.
}
\label{fig:acet_inter_h2o}
\end{center}
\end{figure}

\begin{figure}[!htb]
\begin{center}
\includegraphics[width=0.5\textwidth]{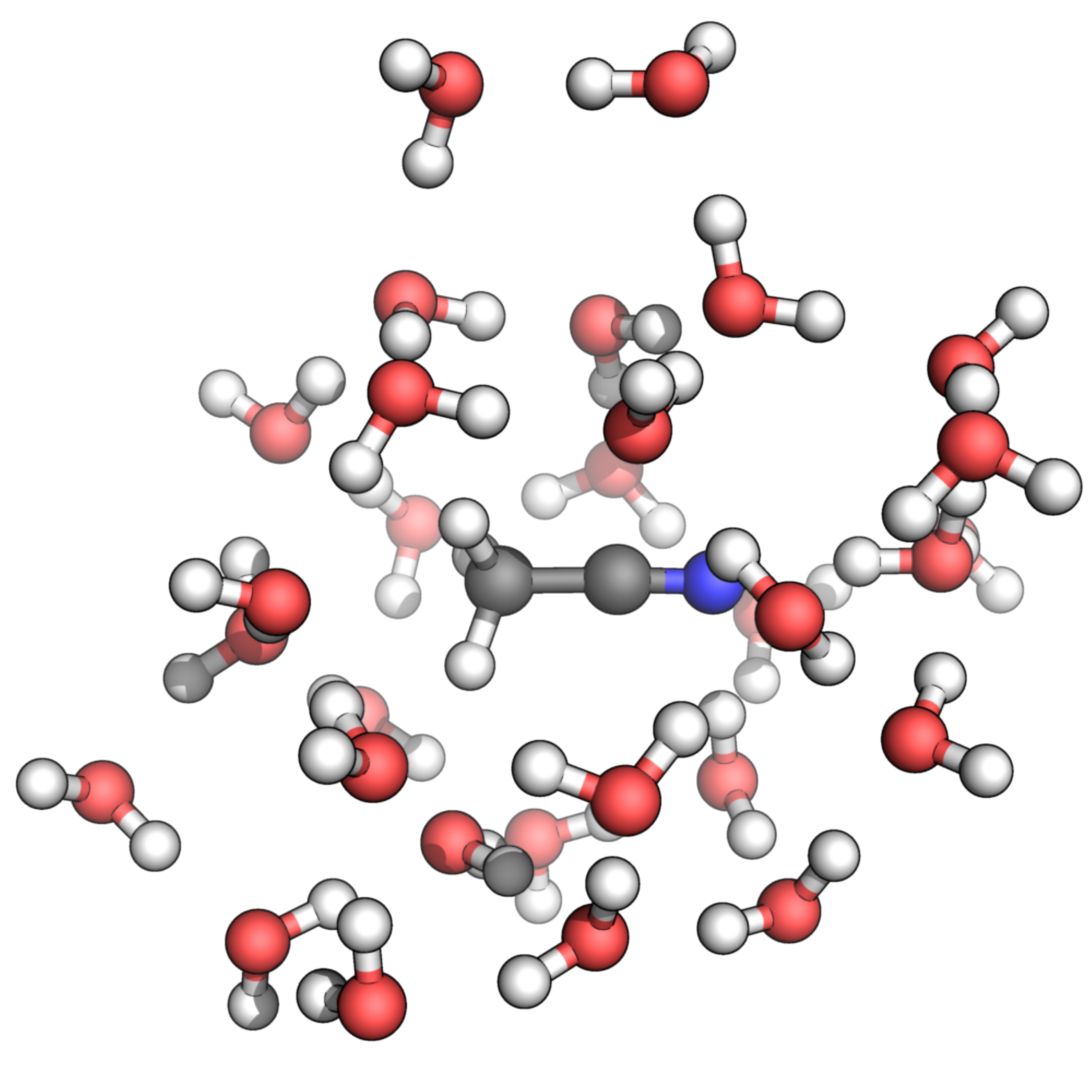}
\caption{
Acetonitrile molecule surrounded by 30 water molecules.
}
\label{fig:h2o_shell}
\end{center}
\end{figure}

In Fig.~\ref{fig:acet_inter_h2o}, $E_\text{inter}$ and the dispersive component of $E_\text{inter}$ (right) of acetonitrile-water complexes (white discs) are plotted as a function of the number of solvent molecules.
First, it can be seen that $E_\text{inter}$ drops from $-30$ to $-140$~kJ/mol with increasing cluster size until the complex consists of 30 molecules;
after that point, $E_\text{inter}$ remains constant (within error bars).
A similar trend can be observed for $E_\text{inter, disp}$.
As expected, more water than DCM molecules are required to reach the point of convergence:
visual inspection of the complexes consisting of 30 water molecules shows that one to two solvation shells are formed (see the most stable configuration in Fig.~\ref{fig:h2o_shell}).
The smaller water molecule interacts more strongly with the solute than DCM ($\langle E_\text{inter} \rangle$ of $-140$ compared to $-90$~kJ/mol).

An outlier for the cluster size of 15 can be identified in Fig.~\ref{fig:acet_inter_h2o}.
This particular cluster (shown in Fig.~\ref{fig:hbonds}) is the only one of that size that features four hydrogen bonds.
As a result, the absolute value of the interaction energy is largest and so is the cluster's weight in the cluster partition function $z$.
However, from Fig.~\ref{fig:hbonds} it can also be seen that a complete solvation shell has not been formed yet.
In Fig.~\ref{fig:acet_inter_h2o}, both $E_\text{inter}$ and $E_\text{inter, disp}$ reflect this fact.

In addition, it can be seen that the spread of $E_\text{inter}$ is large ($\approx 90$~kJ/mol for clusters consisting of 30 solvent molecules).
Considering that all of these structures are minimum energy structures, this finding stresses the importance of rigorous sampling: an inappropriate selection of solvation clusters (from a manual exploration, for instance) can be detrimental to the significance of a theoretical study.

The spread of energies is lower for $E_\text{inter, disp}$ than for $E_\text{inter}$.
Not only is the dispersive component smaller, but dispersive interactions are also
not directional, and hence, not dependent on the exact arrangement of the molecules.
This is in contrast with non-dispersive interactions such as hydrogen bonds that play a central role in this case.

\begin{figure}[!htb]
\begin{center}
\includegraphics[width=0.5\textwidth]{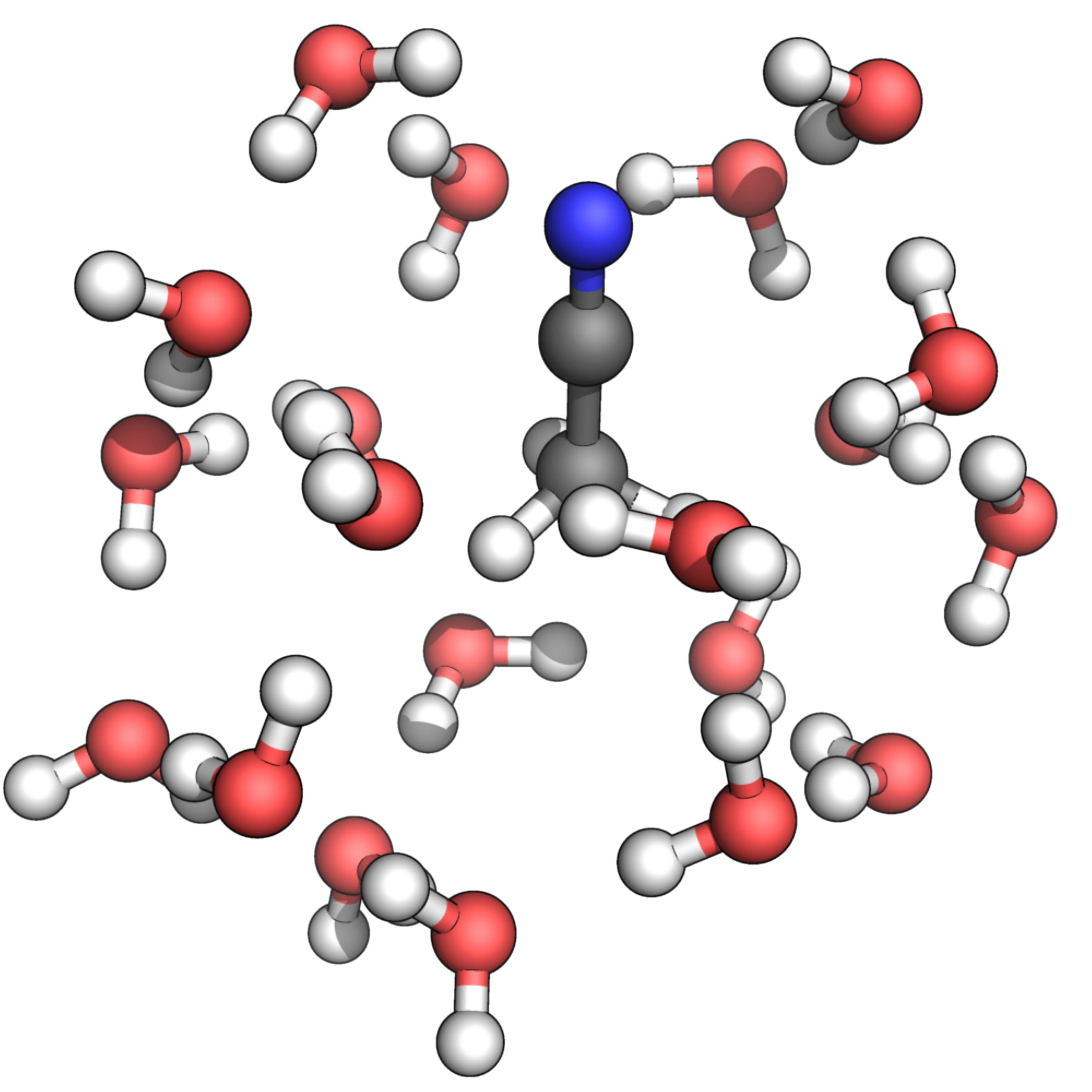}
\caption{
Most stable acetonitrile-water cluster consisting of 20 water molecules.
No hydrogen bonds are formed between the nitrile group and the solvent and the
cluster stabilization is solely brought about by the strong network of solvent-solvent hydrogen bonds.
}
\label{fig:h2o_20}
\end{center}
\end{figure}

Finally, it can be seen that $\langle E_{\text{inter}} \rangle$ and $\langle E_{\text{inter, disp}} \rangle$ do not coincide with $E_{\text{inter}}$ and $E_{\text{inter, disp}}$ of the most stable complexes, respectively.
This difference is more pronounced in water than in DCM (compare Fig.~\ref{fig:acet_inter_dcm}) because in the former the interactions between the solvent molecules are stronger.
In Fig.~\ref{fig:h2o_20}, the most stable acetonitrile-water complex (consisting of 20 solvent molecules) is shown.
This structure is represented by the red point in Fig.~\ref{fig:acet_inter_h2o}, left, for 20 solvent molecules.
In this structure, no hydrogen bond with the acetonitrile group is formed but the water molecules are arranged such that the hydrogen-bond network of the solvent molecules is optimal. However, such a configuration is solely a result of the limited number of
solvent molecules in the solvent shell and would be dissolved in larger clusters, where further water molecules break
up the hydrogen network of the 20 water molecules and allow then for hydrogen bonding with the solute.
However, this most stable configuration with 20 water molecules would have
the highest weight if $E_\text{total}$ had been employed in the Boltzmann weighting.
This emphasizes again how decisive it is to choose a suitable energy measure for calculating the weights $p(\mathbf{x})$.

\begin{figure}[!htb]
\begin{center}
\includegraphics[width=\textwidth]{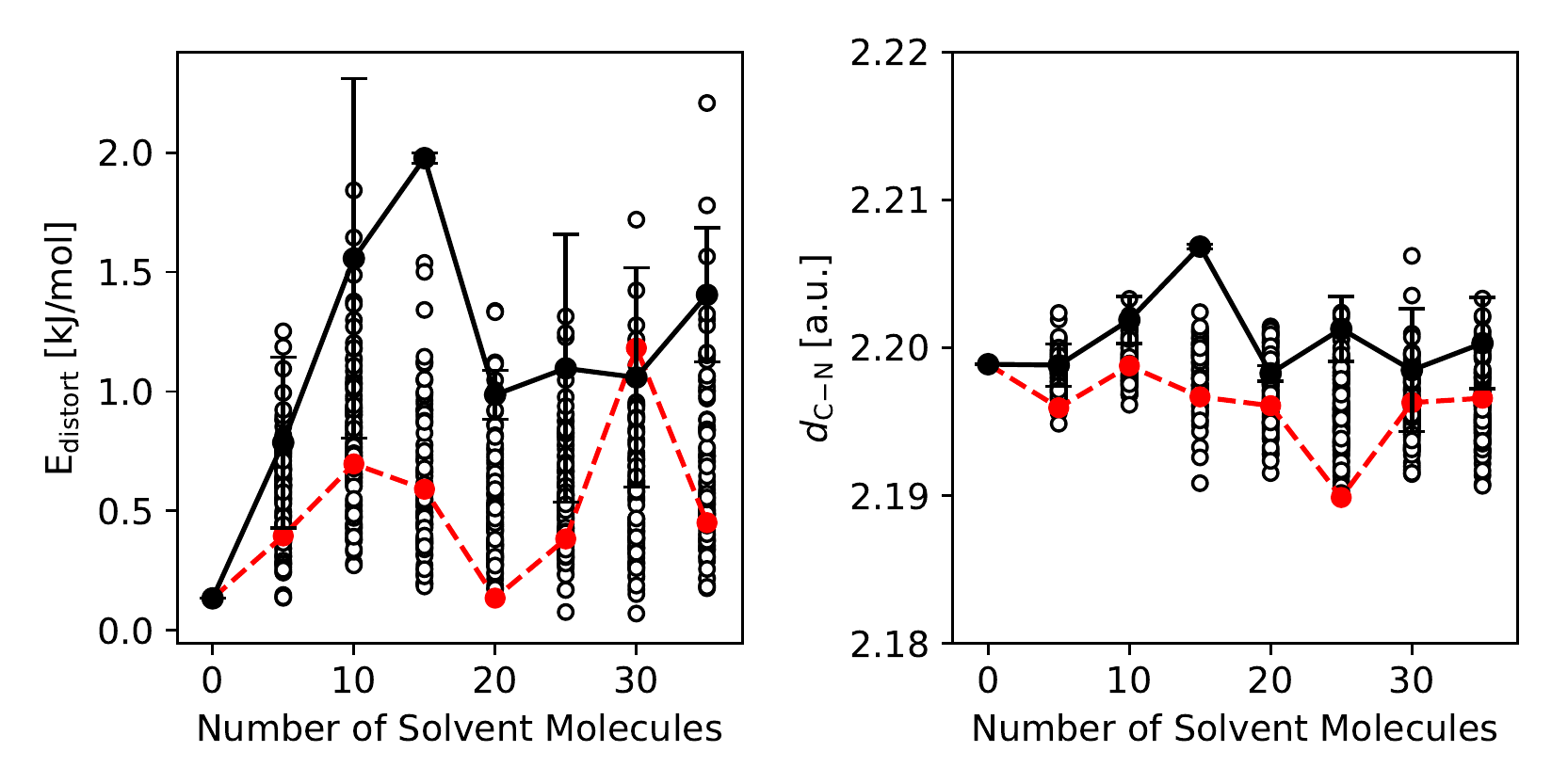}
\caption{
$E_\text{distort}$ (left, in kJ/mol) and $d_\text{C-N}$ (right, in atomic units (a.u.), i.e. bohr) of acetonitrile-water complexes (white discs) as a function of the number of solvent molecules.
Black discs connected by black solid lines show $\langle E_{\text{distort}} \rangle$ and $\langle d_{C-N} \rangle$
(with error bars indicating $\pm 2 \sigma$).
Red discs connected by red dashed lines show $E_{\text{distort}}$ and $d_\text{C-N}$ of the most stable complexes.
}
\label{fig:acet_solute_h2o}
\end{center}
\end{figure}

In Fig.~\ref{fig:acet_solute_h2o}, $E_\text{distort}$ and $d_\text{C-N}$ are shown as a function of the number of water molecules in the solute-solvent complex.
Compared to Fig.~6, it can be seen that $E_\text{distort}$, as well as changes in $d_\text{C-N}$, are larger for all cluster sizes.
The outlier in $\langle E_{\text{inter}} \rangle$ for clusters with 15 water molecules represents the complex shown in Fig.~\ref{fig:hbonds}.
The strong interaction also causes the strong nitrile bond to be elongated.
The presence of additional solvent molecules then reduces the strong solute-solvent interaction.
This is why a similar outlier cannot be found for clusters with more than 15 water molecules.

\begin{figure}[!htb]
\begin{center}
\includegraphics[width=0.5\textwidth]{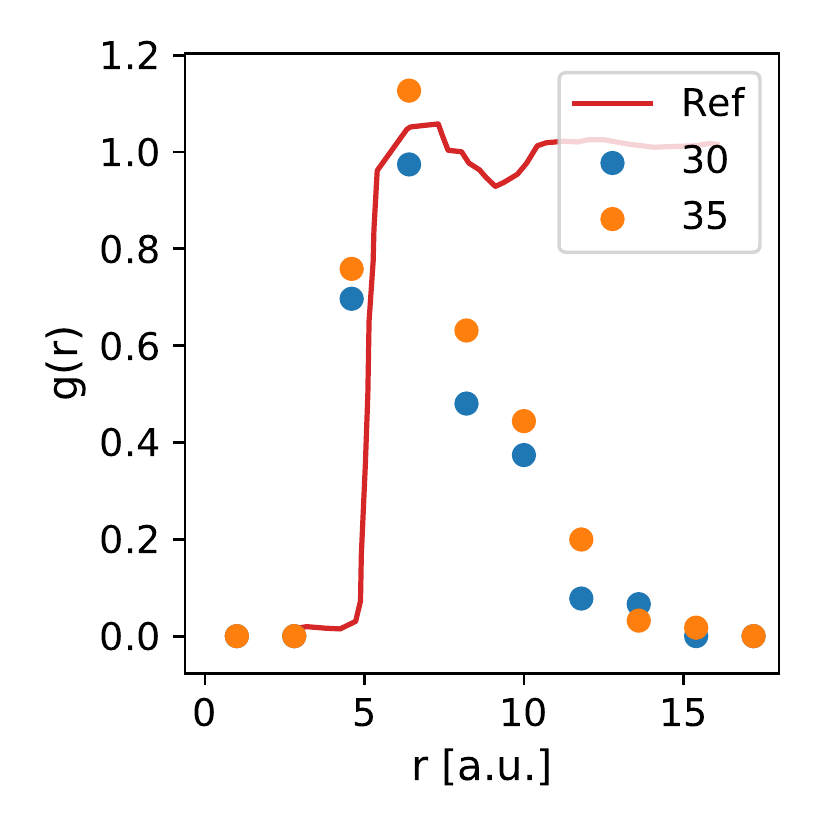}
\caption{Radial distribution function $g(r)$
of the nitrogen-oxygen distance for cluster sizes 30 and 35 (circles)
together with classical molecular dynamics reference data (continuous curve) adapted from Ref. \cite{Bako2005}.
Note the matching position of the maxima.
}
\label{fig:rdf_h2o}
\end{center}
\end{figure}

In Fig.~\ref{fig:rdf_h2o}, the radial distribution function (RDF) $g(r)$ between the nitrogen of acetonitrile and the oxygen of water
for two cluster sizes are shown together with classical molecular dynamics reference data taken from Ref.~\onlinecite{Bako2005}.
For the clusters, the RDF was approximated by taking the weighted average of the RDFs of clusters of the same size.
The cluster sizes 30 and 35 were chosen because clusters of this size describe the
solute-solvent interaction sufficiently well.
It can be seen that clusters' RDFs compare very favorably with the reference data below $r \approx 8$~a.u.
The RDF is close to zero below 5~a.u.\ to reach its maximum at $\approx$ 6.5~a.u.
After that the cluster RDFs drop rapidly as they should because the density of explicit solvent molecules vanishes 
and replaced by implicit solvation of the continuum model in our cluster-continuum approach.
It should be noted that for the calculation of the RDF a relatively large bin size of 1.8~a.u.\
was chosen to ensure that each bin contains sufficiently many solvent molecules.
Further, larger clusters and a larger sample size would be required for a quantitative comparison with simulation data,
especially for the region of the RDF beyond 8 bohr.

\section{Conclusions}

In this study, we presented a systematic approach for modeling solvation effects by establishing a fully automated
static hybrid cluster-continuum scheme with conformational sampling.
Our general algorithm for the stochastic generation of solute-solvent clusters is applicable to any solute or solvent.
We established criteria that not only indicate how many configurations need to be generated
but also how many explicit solvent molecules need to be added to capture most of the interaction between the solute and its environment.
For this, we applied a scheme of weighting sampled configurations to obtain meaningful configuration statistics.
We then proposed measures that indicate how many explicitly treated solvent molecules are necessary to accurately
describe the solute's environment.
As the required size of the clusters depends strongly  on the particular solvent and solute, this
needs to be determined in a case-by-case manner (and in an automated fashion)
when our hybrid cluster-continuum scheme is employed.

We note that the conformational sampling of the solute molecule can be conveniently separated from that
of the solvent shell through, for example, our structure exploration program Chemoton\cite{Simm2017a}, which first
generates solute conformations that can then be solvated. For the sake of efficiency, it is also possible to first
solvate one solute conformation
and then change this conformation in the solvent cages generated to obtain guess structures to be subjected to structure optimization.

The application of our scheme to model acetonitrile in DCM and water highlighted the shortcomings of standard quantum chemical
microsolvation attempts.
The results show, as one would have expected, that in solvents with strong intermolecular interactions (e.g., water)
a single most-stable cluster is unlikely to be representative.
It is not appropriate then to identify only the most stable cluster for calculating ensemble statistics.
A fully automated approach for systematically generating and sampling solute-solvent clusters in a static quantum chemical
picture will be mandatory. The PES of such clusters is too rugged for a rigorous in-depth manual exploration.
This also points toward systematic sampling of configuration space through Monte-Carlo or molecular dynamics algorithms,
to which our automated cluster generation protocol may be seamlessly coupled to eventually allow sampling even under
periodic boundary conditions.

As such extensions will require significant computational resources, a combination
with approximate interaction models ranging from semi-empirical methods to classical molecular-mechanics force fields and
machine learning models is a natural extension (for the fast-growing literature on these schemes, we may refer to references in
Refs.~\onlinecite{Amabilino2019,Lunghi2019,Lu2019,Brunken2019}).
However, we emphasize that reliability of such methods which trade accuracy for computational efficiency is best
guaranteed if suitable uncertainty quantification schemes (such as those reported by us
in Refs.~\onlinecite{Simm2018,Weymuth2018,Proppe2018,Proppe2019})
are in operation that inform about the range of applicability.
Note also that such an approach should then not be considered as
some general transferable model, but as a system-focused baseline model whose suitability is quantitatively assessed by
uncertainty quantification procedures in a rolling fashion through continuous benchmarking as discussed in
Refs.~\onlinecite{Simm2018} and \onlinecite{Proppe2019}.

\section*{Acknowledgments}

This work has been financially supported by the Schweizerischer Nationalfonds (Project No. 200021\textunderscore182400).

\providecommand{\latin}[1]{#1}
\makeatletter
\providecommand{\doi}
  {\begingroup\let\do\@makeother\dospecials
  \catcode`\{=1 \catcode`\}=2 \doi@aux}
\providecommand{\doi@aux}[1]{\endgroup\texttt{#1}}
\makeatother
\providecommand*\mcitethebibliography{\thebibliography}
\csname @ifundefined\endcsname{endmcitethebibliography}
  {\let\endmcitethebibliography\endthebibliography}{}

%
%

\end{document}